SOFTWARE　　　　　　　　　　　　　　　　　　　　　　　　　　　　Open Access

# GWGGI: software for genome-wide gene-gene interaction analysis

Changshuai Wei[1,2] and Qing Lu[1*]

## Abstract

**Background:** While the importance of gene-gene interactions in human diseases has been well recognized, identifying them has been a great challenge, especially through association studies with millions of genetic markers and thousands of individuals. Computationally efficient and powerful tools are in great need for the identification of new gene-gene interactions in high-dimensional association studies.

**Result:** We develop C++ software for genome-wide gene-gene interaction analyses (GWGGI). GWGGI utilizes tree-based algorithms to search a large number of genetic markers for a disease-associated joint association with the consideration of high-order interactions, and then uses non-parametric statistics to test the joint association. The package includes two functions, likelihood ratio Mann–Whitney (LRMW) and Tree Assembling Mann–Whitney (TAMW). We optimize the data storage and computational efficiency of the software, making it feasible to run the genome-wide analysis on a personal computer. The use of GWGGI was demonstrated by using two real data-sets with nearly 500 k genetic markers.

**Conclusion:** Through the empirical study, we demonstrated that the genome-wide gene-gene interaction analysis using GWGGI could be accomplished within a reasonable time on a personal computer (i.e., ~3.5 hours for LRMW and ~10 hours for TAMW). We also showed that LRMW was suitable to detect interaction among a small number of genetic variants with moderate-to-strong marginal effect, while TAMW was useful to detect interaction among a larger number of low-marginal-effect genetic variants.

**Keyword:** Mann–whitney, Non-parametric statistic, Tree model

## Background

The recent genome-wide association studies (GWAS) have made significant progress in finding single genetic variants associated with common complex diseases, with disclosure of regions of interest. Nonetheless, it is likely that a substantial proportion of genetic variants remain uncovered. Common complex diseases are likely caused by the interplay of multiple genetic variants, each with relatively modest marginal effects. A comprehensive genome-wide gene-gene interactions analysis (GWGGI) will take this complexity into account. By fully exploring all available genetic variants on the entire genome, GWGGI could yield novel interaction findings that can be investigated through later bench science research and clinical application for their functional importance and clinical usefulness.

Software, such as the commonly used multifactor dimensionality reduction software (MDR) [1], has been developed for high-dimensional genetic association analyses considering high-order interactions. Nevertheless, only a few statistical packages, such as BOOST [2], Random Jungle [3] and Plink [4], can be directly applied to genome-wide data. In order to detect interactions on a genome-wide scale, researchers commonly use a filter algorithm to pre-select a relative small set of genetic markers prior to the use of gene-gene interactions software (e.g. MDR) or constrain the interaction search to two-way interactions (e.g. Plink). These strategies limit the search space and substantially reduce the computational burden. The trade-off is that important findings, such as high-order interactions and interactions among low-marginal-effect genetic markers, could be missed. To facilitate high-dimensional gene-gene interaction analyses, we developed a C++ package, GWGGI. Similar as several other packages (e.g., MDR), GWGGI evaluates joint association of multiple genetic markers

* Correspondence: qlu@epi.msu.edu
[1]Department of Epidemiology and Biostatistics, Michigan State University, East Lansing, MI 48824, USA
Full list of author information is available at the end of the article considering their possible interactions rather than evaluates interaction-only effects. The package comprises of two modules, the Likelihood Ratio Mann–Whitney (LRMW) module and Tree Assembling Mann–Whitney (TAMW) module. In GWGGI, we further optimize the algorithms and memory storage, making genome-wide analyses feasible on a personal computer with an affordable time.

**Implementation**

The advance of high-throughput technologies generates millions of single-nucleotide polymorphisms (SNPs) that can be extensively explored for gene-gene interactions. Nevertheless, the search of interactions among millions SNPs is a daunting challenge and requires more computationally efficient and powerful statistical tools. We implement two modules, LRMW and TAMW, into GWGGI, for high-dimensional gene-gene interaction analyses involves thousands or even millions of SNPs.

Both LRMW and TAMW utilize the Mann–Whitney U-statistic (MWU) to evaluate joint association of multiple genetic variants with the consideration of possible interactions. Given the multi-locus risk groups formed based on the selected disease-susceptibility genetic variants (the detailed algorithm is described below), LRMW and TAMW assign likelihood ratio (LR) value for each individual belonging to a particular risk group, where LR measures the risk of an individual having disease rather than non-disease [5]. For instance, in LRMW, an individual's LR value is calculated based on his/her genotype of the selected SNPs, $LR_i = \frac{P(G_i|D)}{P(G_i|\bar{D})}$, where $P(G_i|D)$ and $P(G_i|\bar{D})$ are the probabilities of the genotype, $G_i$, in cases and controls, respectively. The MWU can then be formed based on the LR values,

$$U = \frac{1}{N_D N_{\bar{D}}} \sum_{i=1}^{N_D} \sum_{j=1}^{N_{\bar{D}}} \psi(LR_i, LR_j), \quad (1)$$

where $N_D$ and $N_{\bar{D}}$ denote the number of cases and controls, respectively. The kernel function is defined as:

$$\psi(LR_i, LR_j) = \begin{cases} 1, & \text{if } LR_i > LR_j \\ 0.5, & \text{if } LR_i = LR_j \\ 0, & \text{if } LR_i < LR_j \end{cases}.$$

Based on this definition, we can calculate the variance,

$$Var(U) = \frac{1}{N_D(N_D-1)} \sum_{i=1}^{N_D} \left(\frac{1}{N_{\bar{D}}} \sum_{j=1}^{N_{\bar{D}}} \psi(LR_i, LR_j) - U\right)^2 \\ + \frac{1}{N_{\bar{D}}(N_{\bar{D}}-1)} \sum_{j=1}^{N_{\bar{D}}} \left(\frac{1}{N_D} \sum_{i=1}^{N_D} \psi(LR_i, LR_j) - U\right)^2, \quad (2)$$

and build the test statistic, $Z = (U-0.5)/\sqrt{Var(U)}$, which follows a standard normal distribution under null hypothesis. The p-value can thus be calculated to evaluate joint association of identified genetic variants with diseases, considering possible interactions [6,7].

LRMW and TAMW utilize different search algorithms, which make them useful to detect different types of gene-gene interactions. LRMW uses a forward selection algorithm to search all available genetic variants for important interactions among a small number of moderate-marginal-effect genetic variants [7]. In step one, the forward selection algorithm searches all available SNPs for a single SNP to divide samples with different genotypes into two risk groups (e.g., $G_1$={0}, $G_2$={(1, 2)}), which gives the highest possible MWU. In step two, it searches for the second SNP, considering its possible interaction with the first SNP, to split the two existing risk groups into four risk groups (e.g., $G_1$={0, (0, 1)}, $G_2$={0, 2}, $G_3$={(1, 2), (0, 1)}, $G_2$={(1, 2), 2}), which gives the highest possible MWU. The whole splitting process continues until samples are divided into risk groups with a small number of samples. The 10-fold cross-validation is then used to choose the final model with the optimal number of risk groups. As we demonstrate elsewhere [7], the forward selection algorithm is computationally efficient and has the advantage of considering high-order interactions.

For complex diseases influenced by the interplay of hundreds genetic variants, LRMW can only identified those most significant interactions. To consider interactions among hundreds or even thousands genetic variants, most of which have low marginal effects, we also developed TAMW [6]. TAMW uses an ensemble algorithm to combine many de-correlated tree models so as to consider a large ensemble of genetic variants with low marginal effects. In TAMW, a large number of bootstrap samples are first generated from the original data. Each bootstrap sampling creates a bootstrap sample and an out-of-bag (OOB) sample (i.e., individuals left out by the bootstrap sampling). We apply the forward selection algorithm described above to each bootstrap sample, and select a small set of disease-susceptibility genetic variants from a random subset of all genetic variants to build a tree model. The model built from the bootstrap sample is then applied to the corresponding OOB samples by assigning LR values to the individuals in the OOB samples. By repeating the process for all bootstrap samples, we obtain a large number of tree models, each constructed based on a different set of disease-susceptibility genetic variants. The assembling LR for each individual can be calculated by averaging multiple OOB-based LR values, $LR_i^{Assem} = \sum_{j=1}^{T_i} LR_{i,j}/T_i$, where $LR_{i,j}$ is the LR value assigned to an individual $i$ for the $j$-th time, and $T_i$ is the total number of times that the individual $i$ is included in the OOB samples. The overall MWU statistic can then be formed based on the assembling LR to test joint association of hundreds or thousands of genetic variants with diseases. Compared with LRMW, TAMW is more powerful when the underlying disease model involves a large number of genetic variants and their interactions. Nevertheless, it is computationally more intensive than LRMW, and the results from TAMW are less interpretable. Both LRMW and TAMW involve model selection procedure, so the p-value should be obtained by using permutation test. Alternatively, we can split the data to training dataset and testing dataset, and obtain the p-value on testing dataset [6]. Specifically, we first build the model on the training dataset and then apply the model to testing dataset by assigning LR values to the individuals according to their genotypes. A MWU value and the corresponding p-value can be calculated based on the assigned LR values and the case–control outcomes. Note that the MWU value calculated based on equation (1) is also equivalent to AUC [5], which can be used to measure the classification/prediction accuracy of the selected model.

GWGGI is developed in C++, and is available for multiple platforms (e.g., Windows and Linux). The software is based on standard C++ and does not depend on other source codes or libraries. Researchers can download the source code, recompile and build it using different C++ compilers (e.g., Visual Studio in Windows or g++ in Linux or Mac). Figure 1 gives an example of the Windows version of GWGGI.

We implement both LRMW and TAMW into GWGGI, which can be called by using commands "–lmw" and "–tamw", respectively. Users can decide which module to use based on the prior knowledge of the disease and their research purposes. LRMW is suitable for scenarios when the underlying disease model caused by a small number of moderate-marginal-effect genetic variants and interactions, while TAMW fits for disease models involving a large number of low-marginal-effects genetic variants and interactions.

Data read in GWGGI can be either in Plink [4] binary format or in the text format. For data with less than ten thousand SNPs, we suggest the text format, which can be easily read by other statistical software (e.g., R). For data with a much larger size, we recommend Plink binary format, especially for personal computers with limited memory. To optimize the memory storage, the genetic data are stored in memory in a similar way as Plink. Specifically, because two Boolean values can store genetic information for one SNP and one subject, with four possible values, 0, 1, 2, and NA (i.e., the missing value), the genetic information on one locus for all subjects is stored in two Boolean vectors. We use the standard C++ library implementation of vector < bool > for Boolean vectors, which optimizes memory storage so that each value is stored in a single bit. Considering that 1 byte (8 bits) can store 4 SNPs, genetic data with 1 million SNPs for 1000 subjects only needs 250 Mb memory. The tree construction in GWGGI demands most of the computation time. To optimize computation, we use two vectors to store tree information, where one vector is used for storing the location of subjects in the tree structure and the other vector is used for storing

**Figure 1** A screen shot of the GWGGI software.

Table 1 Characteristics of GWGGI

|  | T1D | | CAD | |
|---|---|---|---|---|
|  | LRMW | TAMW | LRMW | TAMW |
| # of SNPs | 2184 | 2184 | 459 K | 459 K |
| # of samples | 4901 | 4901 | 4864 | 4864 |
| Memory usage | 7 Mb | 7 Mb | 731 M | 738 M |
| Loading time | <1 s | <1 s | 3 min | 3 min |
| Analysis time | 1.5 min | 3 min | 3.5 hr | 10 hr |
| Selected SNPs | 7 SNPs | 472 SNPs | 6 SNPs | 57 SNPs |
| AUC | 0.844 | 0.776 | 0.672 | 0.719 |
| P-value* | 1.12e-268 | 3.06e-138 | 1.78e-33 | 4.11e-65 |

*P-values were calculated by applying the model built from the training dataset to the testing dataset.

the LR values of subjects. In this way, the computation can be speed up with little cost of memory use.

GWGGI also provides tuning parameters, so that users can use the software in a flexible and sometimes biologically meaningful way. For example, using "–hz", users can determine whether to include a heterozygote effect (Aa v.s. AA/aa) in the analysis. In TAMW, researchers can also use "–td-burnin" to get the optimal size of trees from the first 50 bootstrap samples.

In the output of GWGGI, both statistical significance of detected gene-gene interactions and the marginal contribution of each genetic variant are reported. In addition, GWGGI provides additional information regarding the joint association model (e.g., providing the AUC value to assess the classification accuracy of the model). More detailed information, such as the tree structure in LRMW and individual LR values in TAMW, can be also obtained if needed.

## Results and Discussion

We demonstrated the use of GWGGI through the analysis of two datasets from the Wellcome Trust genome wide association study [8]. The first dataset comprises of 4901 samples from the Wellcome Trust Type I Diabetes (T1D) GWAS. From the GWAS data, we selected 2184 SNPs that have been previously reported to be associated with T1D or potentially have a function role in T1D. The second dataset is the Genome-Wide dataset of Wellcome Trust Coronary Artery Disease (CAD) study, with 459 k SNPs and 4864 subjects.

The optimal tree size from LRMW (i.e., the highest order interaction considered by LRMW) was chosen based on 10-fold cross validation. The optimal tree size from LRMW was 7 for the T1D dataset and was 6 for the CAD dataset. For TAMW, we chose the optimal size of tree by using the first 50 bootstrap samples (i.e., we first built models on the bootstrap samples, and then chose the optimal tree size based on the overall performance of the models on the OOB samples). The optimal size from TAMW was 4 for the T1D dataset and was 3 for the CAD dataset. The analyses were performed on a personal computer with 2.5GHz CPU and 4G memory. The analysis of the dataset with 2 k SNPs and 4901 samples was completed in a few minutes (Table 1). Through the analysis, LRMW detected a joint association among 6 SNPs in 1.5 minutes, while TAMW identified a joint association among 472 SNPs in 3 minutes. Both joint associations reached statistically significant. From the result of T1D data analysis, we found that the model from LRMW performed slightly better than TAMW. This indicates moderate-to-strong marginal and interactional effects for the T1D-associated SNPs. We also conducted a genome-wide gene-gene interaction analysis by applying GWGGI to the CAD GWAS dataset. For the CAD genome-wide dataset with 459 K SNPs and 4864 subjects, LRMW completed the analysis in 3.5 hours, while TAMW finished the analysis in 10 hours. We observed that TAMW outperformed LRMW in the CAD genome-wide gene-gene interaction analysis. This may indicate that CAD is likely influenced by the interplay

Table 2 Comparison of GWGGI with other software on the T1D dataset

|  | LRMW | TAMW | Plink | BOOST | MDR | RJ |
|---|---|---|---|---|---|---|
| Memory usage | 7 Mb | 7 Mb | 7 Mb | 5 Mb | 56 Mb | 110 Mb |
| Time | 1.5 min | 3 min | 5.5 hr | 14 s | 2 min | 12 min |
| Model | 7 SNPs | 472 SNPs | 2 SNPs | 2 SNPs | 2 SNPs | 1674 SNPs* |
| P-value | 1.12e-268 | 3.06e-138 | 1.74e-30 | 3.31e-62 | 1.27e-20 | 1.27e-103 |
| Selected SNPs** | rs3957146 | rs9273363 | rs9272723 | rs9270986 | rs9273363 | rs9273363 |
|  | rs377763 | rs3957146 | rs9469220 | rs9469220 | rs9275418 | rs3957146 |
|  | rs9270986 | rs9270986 |  |  |  | rs9275523 |
|  | rs9273363 | rs3135377 |  |  |  | rs9275418 |
|  | rs3177928 | rs9275523 |  |  |  | rs9469220 |

*For Random Jungle, we choose the SNPs with permutation importance scores larger than 0.
**SNPs chosen by the methods. If the number is larger than 5, we list the top 5 SNPs.

of a large number of SNPs with low marginal effects. We listed the top SNPs selected by GWGGI in Additional file 1: Table S1.

We also compared GWGGI with several gene-gene interaction analyses packages, including Plink, BOOST, MDR and Random Jungle (RJ). Because not all of the packages were designed for genome-wide analyses, the comparison was made based on the T1D dataset with 2184 SNPs. Among these packages, Plink and BOOST can only consider two-way interactions, while MDR and RJ can explore high-order interactions like GWGGI. Another difference of these packages is how they deal with missing data. BOOST can't handle missing data, therefore we imputed missing data for BOOST. MDR treats missing data as another category besides 0, 1, and 2, while Plink, RJ and GWGGI can directly handle missing data. MDR ran very slowly on a large number of SNPs, so we first applied a "RELIEFF" filter to select top 20 SNPs and then searched for interactions with order up to 5. Because RJ was built for the prediction purpose without providing p-values, we used a goodness-of-fit chi-squared test to calculate p-values. The results were summarized in Table 2. The computational time and the memory usage were calculated based on the training data. The p-values were obtained by evaluating the trained models in the testing data. Among the five methods, BOOST performed best in terms of memory usage (5 Mb) and computational efficiency (14 seconds). Yet, BOOST was limited to non-missing data and two-way interactions. GWGGI was second to BOOST in terms of computational efficiency (7 Mb and 3 minutes). Moreover, GWGGI was able to handle missing data and attained the most significant p-value by exploring high order interactions. Similar as GWGGI, MDR and RJ can also explore high-order interactions. Nevertheless, compared to GWGGI, they required more memory usage and computational time.

## Conclusion

We develop a C++ package, GWGGI, for high-dimensional gene-gene interaction analyses. It comprises of two major functions, TAMW and LRMW, each of which can be used for genome-wide gene-gene interaction analyses without requiring a filter algorithm. In addition, each approach has its own uniqueness. While LRMW is suitable for the identification of gene-gene interactions among a few moderate-marginal-effect genetic variants, TAMW is designed for detecting gene-gene interactions involving hundreds low-marginal-effect genetic variants. We further optimize the package so that the genome-wide gene-gene interaction analysis can be accomplished within a reasonable time on a personal computer.

## Availability and requirement

**Project Name:** GWGGI
**Project Home Page:** https://www.msu.edu/~changs18/software.html#GWGGI or https://www.msu.edu/~qlu/Software.html
**Operating Systems:** UNIX, Windows
**Programming Languages:** C++
**Other Requirements:** No other dependence libraries
**License:** GNU GPL

Any restrictions to use by non-academics: license needed

## Additional file

**Additional file 1: Table S1.** The summary of the top SNPs selected by GWGGI.

**Abbreviations**
GWGGI: Genome-wide gene-gene interaction; LRMW: Likelihood-ratio Mann–Whitney; TAMW: Trees-assembling Mann–Whitney; MWU: Mann–Whitney U; AUC: Area under the curve; T1D: Type 1 diabetes; CAD: Coronary artery disease.

**Competing interests**
The authors declare that they have no competing interests.

**Authors' contributions**
CW and QL participates in the design of the study. CW implemented the methods and drafted the manuscript. QL participated in the conception of the study and in editing the manuscript. Both authors read and approved the final manuscripts.

**Acknowledgements**
This work was supported by the National Institute on Drug Abuse under Award Number K01DA033346 and by the National Institute of Dental & Craniofacial Research under Award Number R03DE022379. This study makes use of data generated by the Wellcome Trust Case Control Consortium.

**Author details**
[1]Department of Epidemiology and Biostatistics, Michigan State University, East Lansing, MI 48824, USA. [2]Department of Biostatistics and Epidemiology, University of North Texas Health Science Center, Fort Worth, TX 76107, USA.

Received: 11 April 2014 Accepted: 11 September 2014
Published online: 16 October 2014

**References**
1. Hahn LW, Ritchie MD, Moore JH: **Multifactor dimensionality reduction software for detecting gene-gene and gene-environment interactions.** *Bioinformatics* 2003, **19**(3):376–382.
2. Wan X, Yang C, Yang Q, Xue H, Fan X, Tang NLS, Yu W: **BOOST: a fast approach to detecting gene-gene interactions in genome-wide case–control studies.** *Am J Hum Genet* 2010, **87**(3):325–340.
3. Schwarz DF, Konig IR, Ziegler A: **On safari to random jungle: a fast implementation of random forests for high-dimensional data.** *Bioinformatics* 2010, **26**(14):1752–1758.
4. Purcell S, Neale B, Todd-Brown K, Thomas L, Ferreira MAR, Bender D, Maller J, Sklar P, De Bakker PIW, Daly MJ, Sham PC: **PLINK: a tool set for whole-genome association and population-based linkage analyses.** *Am J Hum Genet* 2007, **81**(3):559–575.
5. Lu Q, Elston RC: **Using the optimal receiver operating characteristic curve to design a predictive genetic test, exemplified with type 2 diabetes.** *Am J Hum Genet* 2008, **82**(3):641–651.
6. Wei C, Schaid DJ, Lu Q: **Trees assembling Mann–Whitney approach for detecting genome-wide joint association among low-marginal-effect loci.** *Genet Epidemiol* 2013, **37**(1):84–91.

7. Lu Q, Wei C, Ye C, Li M, Elston RC: **A likelihood ratio-based Mann–Whitney approach finds novel replicable joint gene action for type 2 diabetes.** *Genet Epidemiol* 2012, **36**(6):583–593.
8. Burton PR, Clayton DG, Cardon LR, Craddock N, Deloukas P, Duncanson A, Kwiatkowski DP, McCarthy MI, Ouwehand WH, Samani NJ, Todd JA, Donnelly P, Barrett JC, Burton PR, Davison D, Donnelly P, Easton D, Evans D, Leung HT, Marchini JL, Morris AP, Spencer CC, Tobin MD, Cardon LR, Clayton DG, Attwood AP, Boorman JP, Cant B, Everson U, Hussey JM, *et al*: **Genome-wide association study of 14,000 cases of seven common diseases and 3,000 shared controls.** *Nature* 2007, **447**(7145):661–678.